# Verification of the Herschel-Bulkley Model by a Visualization Experiment of a Elastoviscoplastic Yield-Stress Fluid Flow in Straight and Bended Tubes

Mingjun Li[1], Miaorong Du[*,2], Hiroshi Yamaguchi[3], Xiao-Dong Niu[3] and Hiroki Nakakoji[3]

**Abstract**

The elastoviscoplastic yield-stress fluid flows in a horizontal straight tube and a bended tube have been investigated using hydrogen bubble visualization method. The experimental results are used to verify the empirical Herschel-Bulkley model. Both experimental and theoretical investigations well predict the yield-stress fluid flow behaviors. It is found that the significant factors influencing on the predictions of the Herschel-Bulkley model are the yield stress, viscosity and viscoelasticity. For the yield-stress flows in the bended tube, a more delicate constitutive model with consideration of the viscoelastic effects is expected for accurately predicting the flow behaviors.

## I. Introduction

Yield-stress fluids are non-Newtonian materials that flow likes liquids when subjected to a stress or strain above a critical value but responds as elastic or viscoelastic solids below the critical stress or strain (Bird et al. 1982, Barnes 1999). A number of everyday-life and industrial complex fluids ranging from hair gels, cosmetic creams, and toothpastes to many food products

[*] Correspondence author. Email address: dumiaorong@hnu.edu.cn
1． School of Mathematics and Computation Science, Xiangtan University, Xiangtan, 411105, China
2． School of Business Administration, Hunan University, Changsha, 410082, China
3． Energy Conversion Research Center, Doshisha University, Kyoto, 602-0898, Japan

or even concrete and drilling mud exhibit the yield-stress fluid behavior (Bird et al 1982, Moller et al. 2009). Due to the growing use of these non-Newtonian fluids in various manufacturing and processing industries, considerable efforts have been directed towards understanding their flow characteristics (James et al. 1987, Asaga & Roy 1980, Nguyen & Boger 1992, Barnes 1997, 1999, Barnes & Bguyen 2001, Tremblay et al. 2001, Mujumdar et al. 2002, Roussel et al. 2005, Moller et al. 2006).

To model the stress-deformation behavior, several constitutive relations have been proposed and different yield criteria have been used (Ellwood et al. 1990, Barnes 1997, 1999, Benito et al. 2008, Saramito, 2007, 2009). However, there still exist many argues on the constitutive models and the yield stress among several investigators due to the fact of lacking the direct information of these fluids. Most of the yield-stress fluids are non-transparent and difficult to be visualized in the experiment. Therefore, effectiveness of the existent constitutive models is difficult to be validated directly by the actual flow behaviors inside the test devices.

The present study overcomes the above limitation. A transparent elastoviscoplastic yield-stress fluid based on a kind of tooth paste is made for visualization test, and the flow behaviors of it in a horizontal straight tube and a bended tube are investigated, respectively. In the visualization experiment, a hydrogen bubble method (Schraub et al. 1965, and Karweit 1968) is adopted. The hydrogen bubble method can provide very fine bubbles as a tracer in the fluid flow containing water component and those bubbles can thus illustrate the pattern of the water flows in considerable detail. Besides experiment, theoretical and numerical analyses are also performed based on the Herschel-Bulkley constitutive model (Herschel & Bulkey, 1926) for the aim of verifying the model accuracy. The rest of the paper is organized as follows: in section 2 we will discuss the details of the experiment, including test yield-stress fluid, set-ups, etc.; while section 3 focuses on theoretical and numerical analyses. In section 4, experimental and analytical results will be shown and discussed. Finally, a conclusion is given in section 5.

## II. **Experiment**

### *2.1 Test Fluids*

A transparent yield-stress fluid is required for the present visualization test. To this aim, several transparent samples such as toothpaste, paint, clay and printing ink have been investigated, and toothpaste was finally selected as the test fluid because it has consistent rheological characteristics of shear behavior. The test toothpaste is a newly synthesized elastoviscoplastic material with the density of 1,370[kg/m3]. The ingredients of the synthesized toothpaste are listed in Table 1, and its rheological property is shown in Fig. 1, in which the data were obtained from measurement using a rotational Rheometer (Haake RS-75) at room temperature of 20ºC. The results have been shown time independent. As shown in Fig. 1, a non-linear behavior of the shear stress varying with the shear rates denotes the test toothpaste has good elastoviscoplastic feature. The yield stress of the test fluid measured is equal to 99.87[Pa].

### *2.2 Experiment Setup*

Both visualization and measuring tests were carried out in the present experimental investigations. A schematic diagram of experimental apparatus is shown in Fig. 2. The system composes of a syringe pump, which is used to drive the fluid to the test tube, a transparent test section for visualization, a tank for collecting fluid from the test section, and measurement devices for pressure and velocity recording.

In the present study, the yield-stress fluid flows in a horizontal straight tube and a bended tube were investigated, respectively. Figs. 3 (a) and (b) show the dimensions of the test sections of the horizontal straight tube and the bended tube. Measuring points of pressure and velocity are also illustrated in these figures. The tubes in the test section are made of acrylic acid resin for visualization and have diameters of 29[mm].

*2.3    Pressure and velocity profile measurement*

The pressure was measured by using two flush-diaphragm-type pressure transducers (Kyowa: PGM-2KC) which are mounted flush with the test section and directly contacted with the fluid inside.

In order to obtain accurate measurement of the velocity profile of the test fluid flowing through the transparent test section, a hydrogen bubble method (Schraub et al. 1965, and Karweit 1968) was adopted in the visualization experiment. As shown in Fig. 2, a thin platinum wire with diameter of 0.02[mm] as a negative electrode (Cathode) is installed across the tube of the test section. The positive electrode (anode) is made of stainless steel and has a flat shape with area of 100[mm$^2$]. The positive and negative electrodes are connected to a DC power supply device and a pulse signal generator (Microchip: PIC micro 16F84A). The pulse signal produced by the generator has an amplitude 0.1[s] and a period 1[s] (Duty ratio 10[%]), and is amplified by an amplifier. The electric voltage range imposed on the test fluid for the hydrogen bubble generation was 10 ~ 100[V].

When a DC voltage is applied to the wire by the DC power supply device, generating a current that passes through the test fluid, water component in the test fluids electrolyses at the wire surface. The electrolysis produces bubble of hydrogen with diameters comparable to the wire diameter. The bubbles, which are very small, effectively follow the local velocity vector. The traces of the bubbles can be recorded by using a video camera, allowing the velocity profile to be estimated. Comparing other methods like particle tracing method which has to inject smoke, dye or powder into the test fluid, the hydrogen bubble method has merits of no necessary of adding additional tracers in the test fluid (thus not spoiling fluid transparency), less diffusion of hydrogen bubble and easily control by adjusting strength and occurrence time of the electric current.

*2.4  Experimental conditions*

All the experiments in the present study were performed at room temperature 20ºC, and measurements were performed when the test fluid flow in the tubes were considered to be fully developed. For the horizontal straight tube test, an applied voltage imposed across the test section is 100[V/s] and the mass flow rate of the test fluid going into the tube is $6.83 \times 10^{-6}$ [m³/s], which is equivalent to the velocity of 0.0103[m/s]. For the bended tube test, an applied voltage imposed across the test section is 200[V/s] and the mass flow rate of the test fluid going into the tube is $1.65 \times 10^{-6}$ [m³/s], which is equivalent to the velocity of 0.00249[m/s].

### III.  **Theoretical analyses**

With the visualization test, it is possible to examine the accuracy of the existing shear constitutive models of the yield-stress fluids developed in theory by comparing the calculated velocity profile and pressure difference with the experimentally measured data. For elastoviscoplastic materials (e.g., toothpastes), there is no flow until a critical stress (called the yield stress $\tau_0$) is reached. To describe such a material that exhibits a yield stress, the simplest model is the Herschel-Bulkley model (Herschel & Bulkey, 1926) given below:

$$\begin{cases} \dot{\gamma} = 0 & (|\tau| < \tau_0) \\ \tau = \mu|\dot{\gamma}|^{n-1}\dot{\gamma} + \tau_0 & (\tau_0 \geq |\tau|) \end{cases}, \quad (1)$$

where $\tau$ is shear stress, $\dot{\gamma} = du/dr$ the shear rate, and $\mu$ (> 0) the consistency parameter and $n$ (> 0) the power index parameter. Note that $\mu|\gamma|^{n-1}$ has the dimension of a viscosity. When $n = 1$ the model reduces to the Bingham model. The shear thinning behavior is associated with $0 < n < 1$ and the unusual shear thickening behavior to $n > 1$. $\mu$ and $n$ can be obtained by fitting the above equation to the experimental data. In the present study, $\tau_0$, $\mu$ and $n$, by fitting Eq. (1) with the data in Fig. 1, are 99.87[Pa], 45.18[Pa·s] and 0.5824, respectively. Obviously, the test toothpaste has shear thinning property.

*3.1    Analytical solution of the horizontal straight tube*

For axial flow in cylindrical coordinates $(r, z)$, the differential equation for the momentum flux is

$$\frac{d}{dr}(r\tau_{rz}) = \frac{\Delta p}{L} r, \tag{2}$$

where $p$ is pressure and $L$ the length of the tube. On integration of Eq. (2) (with the condition that $\tau_{rz}$ must be finite at $r = 0$), this gives the expression for the shear stress $\tau_{rz}$ for steady laminar flow in a circular tube as:

$$\tau_{rz} = \frac{r}{2}\frac{\Delta p}{L}. \tag{3}$$

Let $r \leq r_0$ be the inner region of the tube where the shear stress is less than the yield stress ($\tau_{rz} \leq \tau_0$) and $r_0 \leq r \leq R$ be the outer region near the tube wall where $\tau_{rz} > \tau_0$. Let $\tau_{rz} = \tau_0$ at $r = r_0$. Then, from Eq. (3), $r_0$ is given by

$$\tau_0 = \frac{r_0}{2}\frac{\Delta p}{L}. \tag{4}$$

Eqs (1), (3) and (4) give the following velocity profile across the tube:

$$u(r) = \int_r^R \left(\frac{\tau - \tau_0}{\mu}\right)^{1/n} dr. \tag{5}$$

In the situation of $\tau < \tau_0$, Eq. (5) gives:

$$u(r) = \frac{nR}{n+1}\left(\frac{\Delta pR}{2L\mu}\right)^{\frac{1}{n}}\left(1 - \frac{2L\tau_0}{\Delta pR}\right)^{\frac{1+n}{n}}, \tag{6}$$

and in the situation of $\tau_0 < \tau$, Eq. (5) gives:

$$u(r) = \frac{nR}{n+1}\left(\frac{\Delta pR}{2L\mu}\right)^{\frac{1}{n}}\left[\left(1 - \frac{2L\tau_0}{\Delta pR}\right)^{\frac{1+n}{n}} - \left(\frac{r}{R} - \frac{2L\tau_0}{\Delta pR}\right)^{\frac{1+n}{n}}\right]. \tag{7}$$

Eq. (6) implies the fluid is in plug flow in the center region of the tube where $\tau < \tau_0$ and Eq. (7) denotes that the velocity profile is parabolic in the region near the tube wall where $\tau_0 < \tau$.

The theoretical pressure difference $\Delta p$ in Eqs. (6) and (7) is given by the mass flow rate. With Eqs. (6) and (7), we have

$$Q = \int_0^R \pi r^2 \frac{du}{dr} dr = \frac{\pi R^3}{\mu^{1/n}}\left[\left(\frac{\Delta pR}{2L} - \tau_0\right)\right]^{1/n}$$

$$\times \left[\frac{n}{3n+1}\left(1 - \frac{2\tau_0 L}{\Delta pR}\right)^3 + \frac{2n}{2n+1}\left(1 - \frac{2\tau_0 L}{\Delta pR}\right)^2 \frac{2\tau_0 L}{\Delta pR} + \frac{n}{n+1}\left(1 - \frac{2\tau_0 L}{\Delta pR}\right)\left(\frac{2\tau_0 L}{\Delta pR}\right)^2\right] \tag{8}$$

*3.2    Solution of the bended tube*

Unlike the straight tube flow, analytical solution of the bended tube flow is difficult to obtain. Therefore, in the present study, we used a CFD software FLUENT based on the SIMPLE (Semi-Implicit Method for Pressure Linked Equations) method to investigate the numerical solution of the bended tube flow. In brief, the following governing equations on a non-uniform grid were solved based on the conditions of (inlet: fully developed velocity; outlet: constant pressure and zero flux; wall: non-slip):

$$\nabla \mathbf{u} = 0, \tag{9}$$

$$\frac{D(\rho \mathbf{u})}{Dt} = -\nabla P + \nabla[\eta(\dot{\gamma})\dot{\gamma}] + \rho g, \tag{10}$$

where $\eta(\dot{\gamma})$ is the general viscosity given from Eq. (1) as

$$\eta(\dot{\gamma}) = \mu\|\dot{\gamma}\|^{n-1} + \frac{\tau_0}{\sqrt{2}\|\dot{\gamma}\|}. \tag{11}$$

IV. **Results and Discussions**

*4.1 Horizontal straight tube*

Fig. 4 shows the visualization of the flow pattern of the test toothpaste fluid in the horizontal straight tube. Typical plugging velocity profiles in the center region of the tube can be clearly observed along the test tube. Near the wall, where the stress is expected to be less than the yield stress, a parabolic shape of the velocity is seen. Comparisons of the experimental data and the theoretical solutions of Eqs. (6), (7) and (8) are given in Fig. 5 and Table 2, respectively. Fig. 5 shows the comparison of the experimentally measured velocity profiles of Fig. 4 and the theoretical solution and Table 6 presents the comparisons of the radius of plugging velocity (flat part of the velocity profile in Fig. 5), maximum velocity and pressure difference along the test tube.

As shown in Fig. 5, the theoretical solution based on the Herschel-Bulkley model (Eq. (1)) agrees well with the experimental data near the wall, and there is a maximum difference of 2.3% between two results in the center region. The pressure difference predicted by the theoretical solution is 34.08[kPa], which is 13.6% larger than that of experimental measurement. We attributed the above differences of the velocity and pressure difference to the simplification of the theoretical analysis, particularly the empirical based Herschel-Bulkley model of Eq. (1) in the inner region of the tube ($\tau < \tau_0$). However, the above errors are acceptable and the Herschel-Bulkley model of Eq. (1) holds a reasonable accuracy for the test toothpaste fluid.

As seen from Eqs. (1) and (8), the influence factors to the above errors mainly are from three model parameters of $n$, $\mu$, $\tau_0$, and the mass flow rate $Q$. To exactly see the effects of these factors ($n$, $\mu$, $\tau_0$, and $Q$) to the plugging radius, the maximum velocity and the pressure difference, we introduce a relative uncertainty of 5% to one of the four influence factors and fixes others, and then calculate the deviation of the plugging radius, the maximum velocity and the pressure difference, respectively. Table 3 shows the calculated the deviations of the plugging radius, the maximum velocity and the pressure difference to the influence factors of $n$, $\mu$, $\tau_0$, and $Q$, respectively. As shown in Table 3, it is found that the yield stress $\tau_0$ and the viscosity $\mu$ are the most significant influence factors to the plugging radius and the pressure difference, while the mass flow rate $Q$ is mostly influencing the velocity but with very small amplitude. Therefore, accurately expressing the yield stress and the viscosity is important to improve the accuracy of the Herschel-Bulkley model.

## *4.2  Bended tube*

The visualizations of the flow pattern of the test yield-stress fluid in the horizontal inlet and vertical outlet parts of the bended tube are shown in Fig. 6, respectively. In the horizontal inlet part, typical plugging velocity profiles in the center region are again clearly observed along the test section, and in the vertical part of the test section, an asymmetric velocity profiles are detected. Near the wall, a parabolic shape of the velocities is also observed for the flows in both the inlet and outlet parts of the test tube. Comparisons of the experimentally measured velocity profiles of Fig. 6 and the calculation results of Eqs. (9), (10) and (11) are given in Fig. 7 for the flows in the inlet and outlet parts of the test section, respectively. The calculation results are found relatively well agreement with the experimental data, and successfully predict the plugging and asymmetric velocity profiles in the respective inlet and outlet parts of test section. Due to the over simplification of the Herschel-Bulkley model (Eq. (1)), the maximum errors between two results in the inlet and outlet tubes are found to be 19% and 9%, respectively. The larger error between the theoretical calculation and the experimental results in the inlet tube is chiefly from the viscoelastic property of the test fluid, which makes the flow pattern in the vertically bended tube more complicated as compared with the simple shear mode (Saramito, 2007, 2009).

The pressure difference has a larger discrepancy between the experimental measurement, 11[kPa] and the theoretical calculation, 51[kPa] in the bended tube test than that in the horizontal straight tube experiment. This also probably attributes to the lacking consideration of the viscoelasticiy of the test fluid in the Herschel-Bulkley model (Saramito, 2007, 2009). It is thought, the effects of the viscoelasticity are constant everywhere in the horizontal straight tube test as observed from the velocity profile visulaization in Fig. 4, and have no contribution to the pressure. Therefore, there is negligible difference of the pressure losses between the measurement and calculation. However, for the elastoviscoplastic toothpaste flow in the bended tube, particularly in the vertically bended part of it, the effects of the viscoelaticity to the pressure are expected larger than those in the horizontal tube test, leading to the pressure loss decrease in the outlet of the tube. Therefore, the measured pressure difference is much less than the calculated value for the bended tube.

## V.     Conclusions

The elastoviscoplastic yield-stress flows in a horizontal straight tube and a bended tube were investigated through an experimental visualization. The experimental results are used to validate the simple empirical Herschel-Bulkley model. Based on experimental data and theoretical analyses, we have the following conclusions:

(1) Both experimental and theoretical investigations well predict the yield-stress flow behaviors (e.g., plug flow) in the horizontal straight tube and the bended tube.

(2) The significant factors influencing on the theoretical predictions are the yield stress and the viscosity.

(3) For the elastoviscoplastic flows in a complex container like the bended tube, a more delicate constitutive model with consideration viscoelasticity is expected for improving the theoretical prediction of the flow behaviors.

**Table 1: Ingredients of test toothpaste**

| Ingredients | Name |
|---|---|
| Moistening agent | Sorbitol |
| Cleaning agent | Anhydrous silicic acid |
| Dissolving agent | Polyoxyethylen alkyl sulfosuccinate |
| Medical component | Polyethyleneglycol, Sodium fluoride, Isopropyl methyl phenol |
| Flavoring agent | Aroma chemical (Saccharine sodium) |
| Bloating agent | Sodium lauryl sulfate |
| Bond | Xanthan gum |
| Firming agent | Oxidized titanium |

**Table 2: Comparison of experimental data and theoretical solution of the toothpaste flow in the straight tube**

| | Experiment | Theoretical solution |
|---|---|---|
| Plug radius ($r_0/R$) | 0.49 | 0.48 |
| Maximum velocity ($u_{max}/U_{average}$) | 1.46 | 1.49 |
| Pressure difference $\Delta p$ [kPa] | 30 | 34.08 |

**Table 3: Error analyses of the Herschel-Bulkley model for the straight tube flow**

| Influence factor | $\tau_0$ | n | $\mu$ | $Q$ |
|---|---|---|---|---|
| Deviation of plug radius ($r_0/R$) | 0.021 | 0.013 | 0.021 | 0.01 |
| Deviation of maximum velocity ($u_{max}/U_{average}$) | $1.33 \times 10^{-4}$ | $2.59 \times 10^{-4}$ | $1.35 \times 10^{-4}$ | $12.8 \times 10^{-4}$ |
| Deviation of pressure difference $\Delta p$ [kPa] | 1.628 | 0.738 | 1.246 | 0.726 |

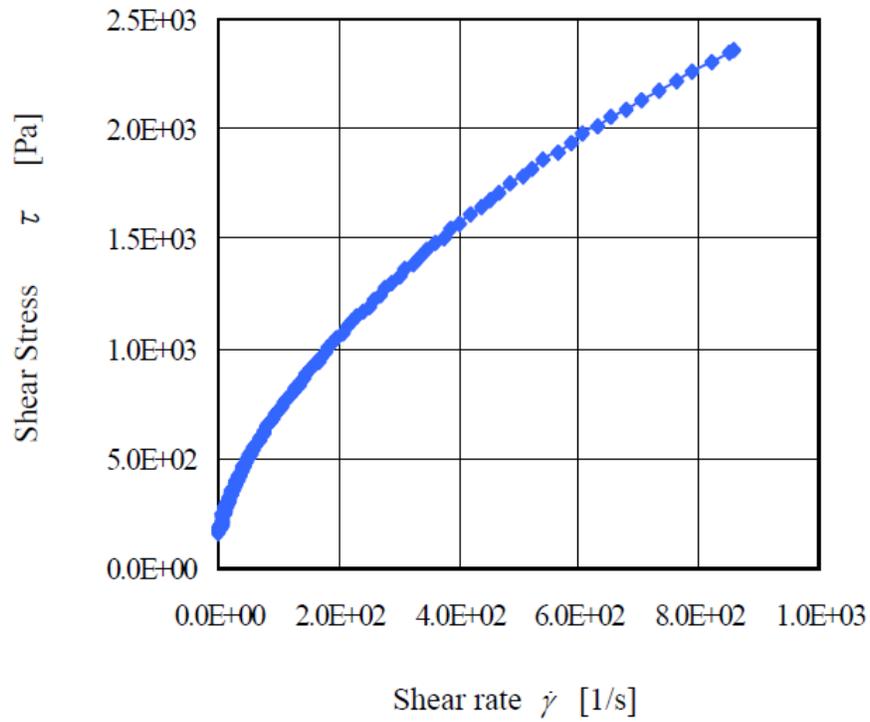

Fig. 1: Rheological property of test fluid

① Pump
② Pressure sensor
③ Test section (Acryl)
④ Tank
⑤ DC power supply
⑥ Ampere meter
⑦ Pulse signal generator
⑧ Electric pole (SUS)

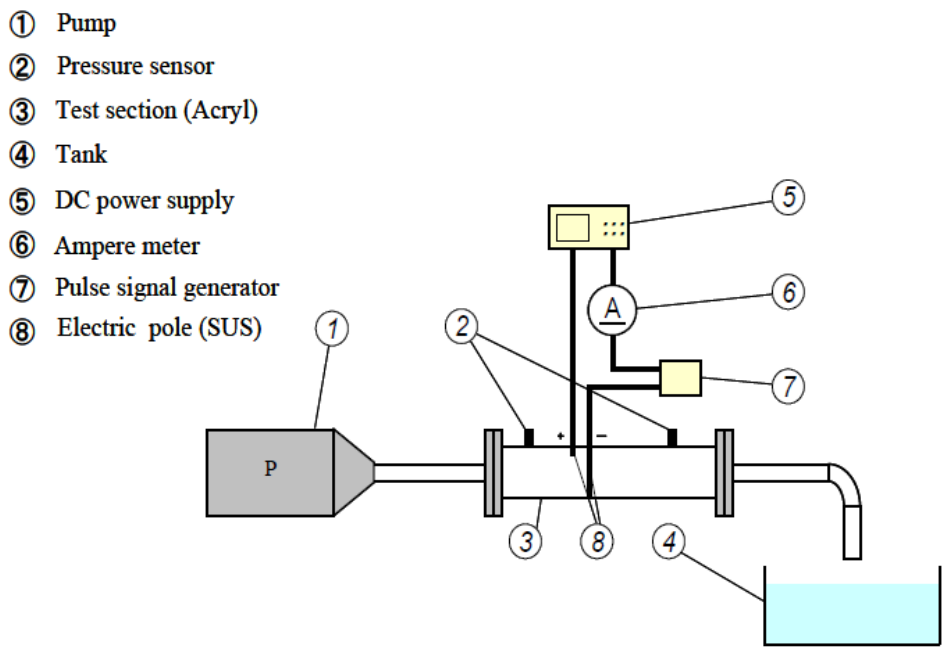

Fig. 2: Schematic diagram of experimental set-up

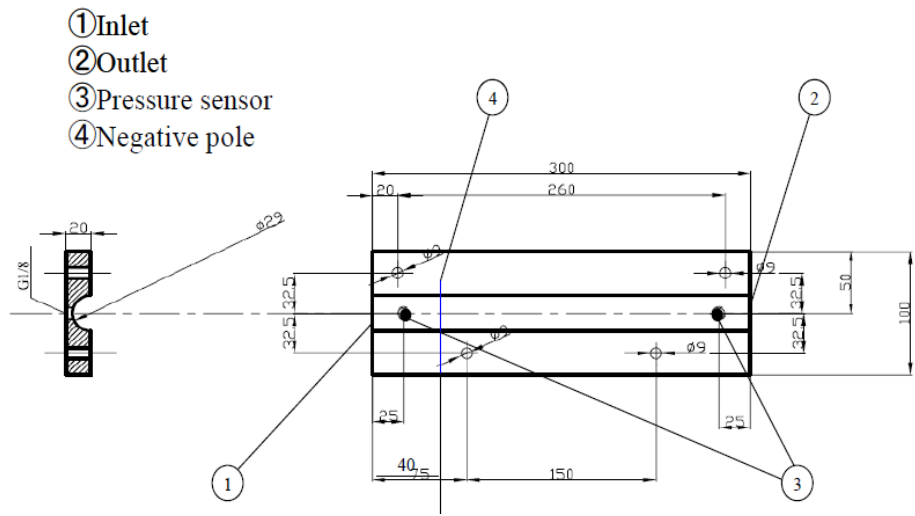

(a) Straight tube

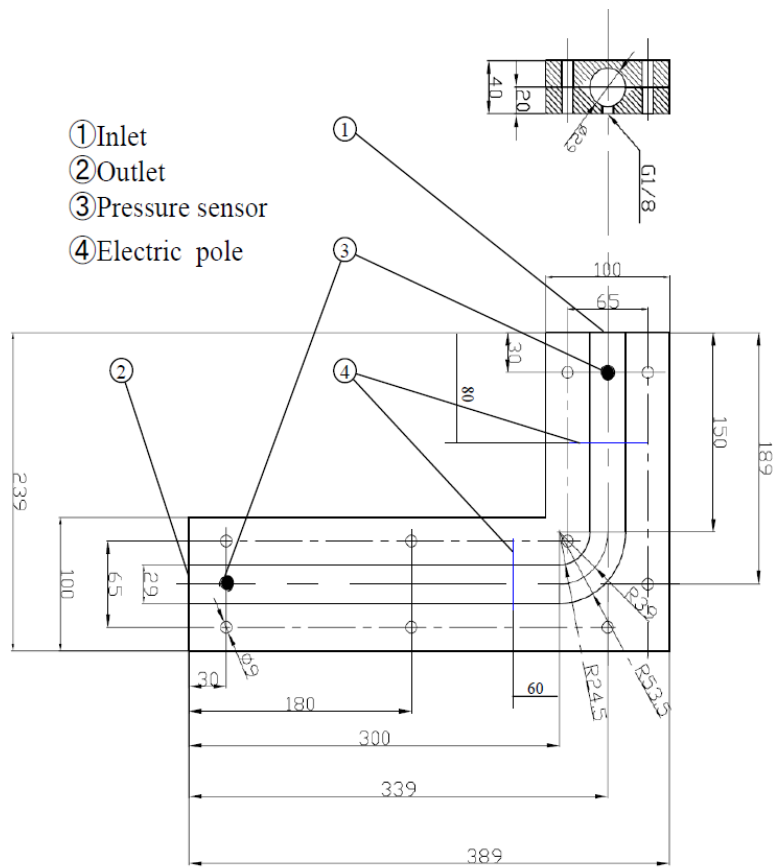

(b) Bended tube

Fig. 3: Dimensions of test sections and measuring positions along them

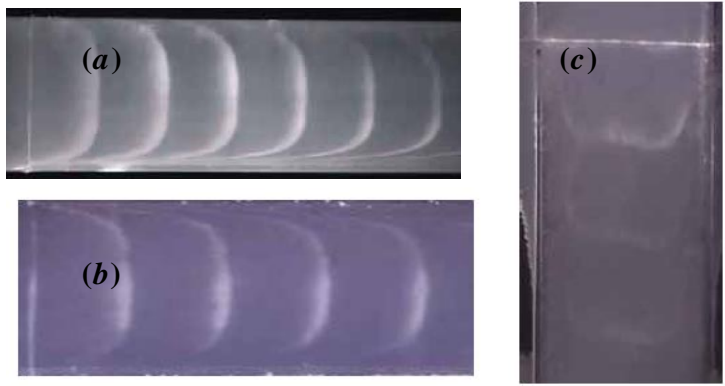

Fig. 4: (a)Visualization of velocity profile in the straight tube

Fig. 5: Visualization of velocity profile in the bended tube (Left: (b) Horizontal inlet part; Right: (c) Vertical outlet part)

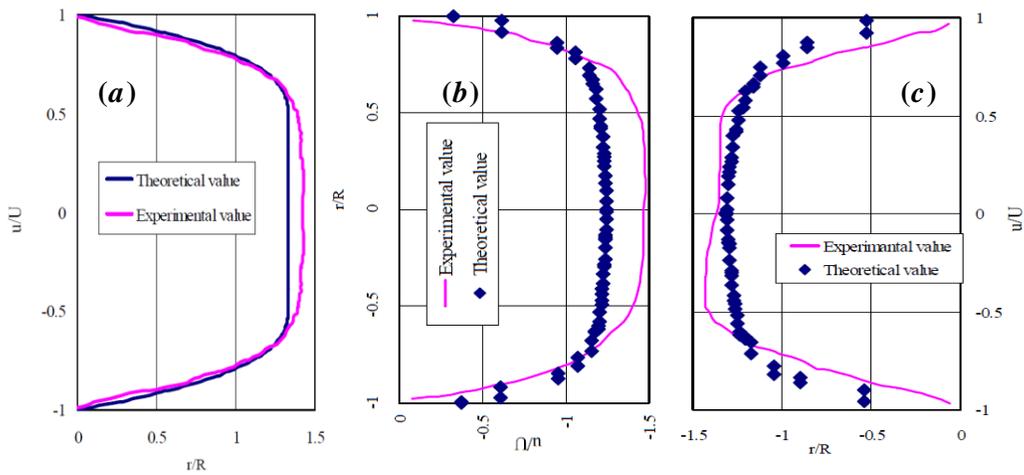

Fig. 6: (a) Comparison of velocity profiles obtained by measurement and theoretical analysis for the straight tube

Fig. 7: Comparison of velocity profiles obtained by PIV and theoretical solution for the bended tube (Left: (b) Horizontal inlet part; Right: (c) Vertical outlet part)